\pdfoutput=1
\documentclass[letterpaper]{sig-alternate-05-2015}
\usepackage{times}
\usepackage{helvet}
\usepackage{courier}
\usepackage{graphicx}
\usepackage{caption}
\usepackage{subcaption}
\usepackage{amsmath,amssymb}
\usepackage{url}
\usepackage{MnSymbol} 
\usepackage{multirow}
\usepackage{etoolbox}
\makeatletter
\patchcmd{\maketitle}{\@copyrightspace}{}{}{}
\makeatother
\usepackage[aboveskip=1pt,labelfont=bf,labelsep=period,justification=raggedright,singlelinecheck=off]{caption}
\usepackage[justification=centering]{caption}

\date{}

\begin{document}

\title{The International Postal Network and Other Global Flows As Proxies for National Wellbeing}
\numberofauthors{5}

\author{
\alignauthor
  Desislava Hristova\bf{*}\\
\affaddr{Computer Laboratory\\
 University of Cambridge, UK}
\alignauthor
  Alex Rutherford\\
  \affaddr{Global Pulse, United Nations\\ 
  New York, USA}
 \and
\alignauthor
  Jose Anson\\
  \affaddr{Universal Postal Union\\ 
  Bern, Switzerland}
\alignauthor
  Miguel Luengo-Oroz\\
  \affaddr{Global Pulse, United Nations\\ 
  New York, USA}
\alignauthor
  Cecilia Mascolo\\
\affaddr{Computer Laboratory\\
 University of Cambridge, UK}
}
\maketitle

\let\thefootnote\relax\footnote{\textbf{*}Corresponding author: desislava.hristova@cl.cam.ac.uk}

\begin{abstract}
The digital exhaust left by flows of physical and digital commodities provides a rich measure of the nature, strength and significance of relationships between countries in the global network. With this work, we examine how these traces and the network structure can reveal the socioeconomic profile of different countries. We take into account multiple international networks of physical and digital flows, including the previously unexplored international postal network. By measuring the position of each country in the Trade, Postal, Migration, International Flights, IP and Digital Communications networks, we are able to build proxies for a number of crucial socioeconomic indicators such as GDP per capita and the Human Development Index ranking along with twelve other indicators used as benchmarks of national well-being by the United Nations and other international organisations. In this context, we have also proposed and evaluated a global connectivity degree measure applying multiplex theory across the six networks that accounts for the strength of relationships between countries. We conclude with a multiplex community analysis of the global flow networks, showing how countries with shared community membership over multiple networks have similar socioeconomic profiles. Combining multiple flow data sources into global multiplex networks can help understand the forces which drive economic activity on a global level. Such an ability to infer such proxy indicators in a context of incomplete information is extremely timely in light of recent discussions on measurement of indicators relevant to the Sustainable Development Goals. 
\end{abstract}

\section{Introduction}
The vast streams of data that are produced by the use of automated digital services such as social media, email and mobile phones, also known as `Big Data', have for some time been leveraged in the private sector to assist in tasks as diverse as logistics, targeted advertising and offering personalised multimedia content. More recently, these same data sources and methodologies have begun to be used to assist humanitarian and development organisations, allowing new ways to use data to implement, monitor and evaluate programs and policies~\cite{ungp}. The ability of such novel data sources to complement traditional data collection techniques such as household surveys and focus groups is clear~\cite{ungp2}. The data is collected passively without the need for costly and potentially dangerous active data collection, which also avoids inaccuracies due to human error, bias~\cite{arnulf} or dishonesty.

However, the use of Big Data for development is still relatively nascent and questions remain over the ability of such sources to measure or approximate metrics of interest. Invariably, data sources such as social networking applications enjoy deeper penetration in developed economies and rely on expensive technologies such as smart phones and robust communications infrastructure. It has been noted that measurements of human dynamics based on such recent platforms can lead to strong biases~\cite{tufekci}, with worse implications for those with limited access to these digital platforms.

In this paper we present analysis of a data source which is undoubtedly `Big' yet represents one of the most established and pervasive long-distance communications networks in the history of mankind. The international postal network (IPN) established in 1874 is administered by a dedicated United Nations specialised agency: the Universal Postal Union (UPU). Due to regulatory reporting requirements and the capabilities of automated data capture technologies such as RFID tags, the records of individual postal items maintained by UPU represent a rich record of human activity with unparalleled penetration, which can be expected to reflect individual level behaviour, local, regional and national economic activity and international economic relations.

Network representations have emerged as an extremely powerful and general framework for analysing and modeling systems as diverse as transportation, biological processes, academic authorship and logistics among others~\cite{barabasi13}. Network science provides powerful tools for understanding such systems with large sets of coupled components with emergent behaviours more generally known as complex systems. Previous work has explored flows of both physical and digital nature, where physical flows of goods and people~\cite{trademulti,networkorigins,culturalhistory,cargo,productspace,airline,fingerprints} and digital flows of information and communication~\cite{tag,eagle2010,ugander,weber,bogdan2015} have been extensively studied in the past in order to understand better the way in which they affect the wealth, resilience and function of social systems on global, regional, national and sub-national scales. With our work we aim to address the general question of whether \emph{structural network properties of different flow networks between countries can be used to produce proxy indicators for the socioeconomic profile of a country}.

\subsection*{Global Multiplexity}
A natural extension of a network in which edges between pairs of nodes represent a single kind of flow between those nodes, is to a \textit{multiplex} model~\cite{multiplex} including several qualitatively different kinds of flows which may each be understood as a single distinct layer. The advantages of a multiplex model is that the aggregation of several different network layers have been shown to be more informative than a single layer~\cite{sampling,trademulti}. This is particularly true if some layers are partially or wholly unobserved or if one layer imposes a barrier to entry in the form of a cost for an edge to form. For example, flows of trade and flights rely on bilateral agreements and legal frameworks as well as predictable demand to be commercially viable. In contrast, personal communications flows such as email are more readily initiated, requiring only that at least one participant has the email addresses of all the others. 

Multiplexity, or the multiple layers of interactions between the same entities, has been explored in a wide range of systems from global air transportation~\cite{cardillo2013} to massive online multiplayer games~\cite{Szell19072010}. In~\cite{hay2005}, the author studied the implications of multiple media usage on social ties in an academic organisation and discovered that multiplex ties (those which use multiple media) indicate a stronger bond. This has been empirically evaluated on networks with both geographical and social interactions recently~\cite{me2014}, where it was found that people share a stronger bond when observed to communicate through many different media. These findings support the intuition that a pair of nodes enjoy a stronger relationship if they are better connected across several diverse network layers. The multichannel exchange of information or goods, offers a simple and reliable way of estimating tie strength but has not been applied to international networks of flows until now.

\section{Methodology and Data}
In this work, we explore over four years of daily postal data records between 187 countries by comparing it to other global flow networks. Those of a physical nature, such as the trade, migration and digital networks. We show how the network properties of global flow networks can approximate critical socioeconomic indicators and how network communities formed across physical and digital flow networks can reveal socioeconomic similarities possibly indicating dependencies within clusters of countries. 

Real-time measurements of international flow networks can ultimately act as global monitors of wellbeing with positive implications for international development efforts.
Using knowledge about the way in which countries interact through flows of goods, people and information, we use the principles of multiplexity theory to understand the strength of international ties and the network communities they form. In this section, we will detail the methods used to perform our analysis and the various datasets with focus on the international postal network (IPN), which has previously not been described. 

\subsection{Multiplex model} A comprehensive review of multiplex network models can be found in~\cite{kivela2013}, however, in this work we will apply a simple multiplex model to capture the multiple flow interactions which we will describe in the following section. In our model, we consider all six networks in our study as a collection of graphs: 
\begin{equation}
{\cal{M}}= \{G^1(V^1,E^1),...,G^\alpha(V^\alpha,E^\alpha),...,G^m(V^m,E^m)\}
\end{equation}

where each graph contains a set of edges $E$ and nodes $V$, and $m$ is the total number of networks. This allows us to define the multiplex neighbourhood of a node $i$ as the union of its neighbourhoods on each single graph: 

\begin{equation}
N_{\cal{M}}(i) = \{N_\alpha(i) \bigcup N_\beta(i) ... \bigcup N_m(i)\}
\end{equation}

where $N_\alpha(i)$ is the neighbourhood of nodes to which node $i$ is connected on layer $\alpha$.
The cardinality of this set can be considered as the node's global multiplex degree, or in other words the number of countries with which a country has exchanges:

\begin{equation}
k^{glob}(i) = |N_{\cal{M}}(i)|
\end{equation}

From the multiplex neighbourhood, we can also compute the weighted global degree of a node $i$ as:

\begin{equation}
k^{glob}_w(i) = \frac{\displaystyle\sum_{j \in N_{\cal{M}}(i)}\displaystyle\sum_{G \in \cal{M}} e_{ji}}{n *m}
\end{equation} 

\begin{figure}[t!]
\centering
 \begin{subfigure}[b]{0.25\textwidth}
                \includegraphics[width=\textwidth]{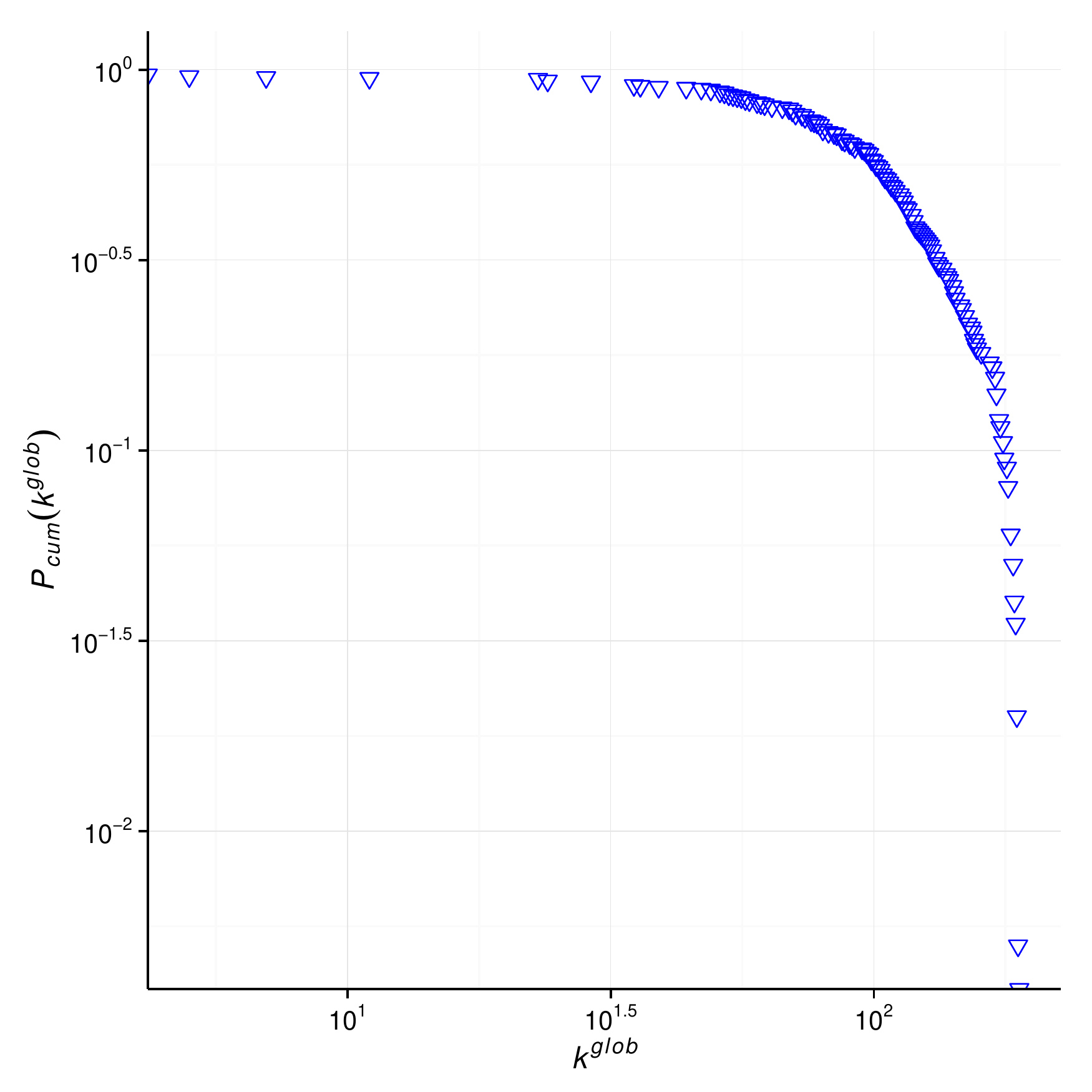}
\end{subfigure}%
           \begin{subfigure}[b]{0.25\textwidth}
                \includegraphics[width=\textwidth]{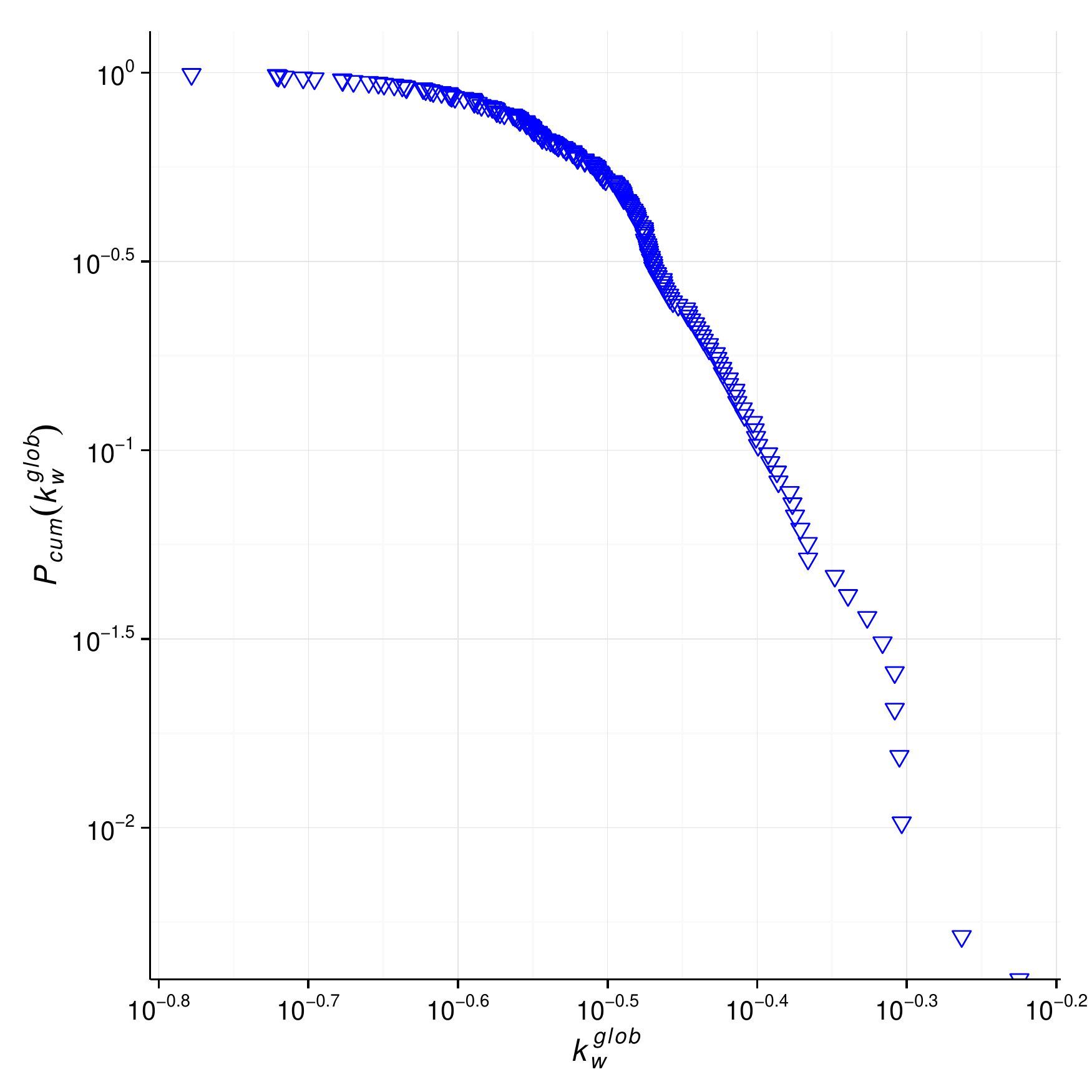}
\end{subfigure}%
\caption{CCDF of weighted and unweighted global multiplex degrees.}
\label{fig:gdegs}
\end{figure}

This is the sum of the weights of edges in the multiplex neighbourhood and for each graph layer they appear on. We add an edge weight if $e_{ij}, e_{ji} \in G$ for each network in the collection ${\cal{M}}$. We only consider edges present in both directions because the global degree is ultimately a measure of tie strength and we want to consider well-established flows between countries only. This is common practice in other contexts where tie strength is of importance such as in social networks~\cite{kwak}. We then normalise the weighted global degree by the number of possible edges $n*m$, where $n$ is the total number of nodes and $m$ is the number of networks in the multiplex collection.  We plot the cumulative degree distribution of both the weighted and unweighted global degrees in Fig.~\ref{fig:gdegs}. 

\subsection{Community multiplexity} Networks are powerful representations of complex systems with a large degree of interdependence. However in many such systems, the network representing it naturally partitions into communities made up of nodes that share dependencies between each other, but share fewer with other components. In the present context, communities are composed of groups of countries that share higher connectivity than the rest of the network.  If two countries appear in the same community in most of the six networks, this can be considered a greater level of interconnectedness otherwise not visible from the single network perspective. We formalise this idea as the \emph{community multiplexity} of a pair of countries $(i,j)$:

\begin{equation}
cm(i,j) = \displaystyle\sum_{G \in M}^m \delta(c_i^G, c_j^G)
\end{equation}

where $c_i$ is a discrete variable indexing the cluster of which country $i$ is a member. If the two are equivalent for a given network $G$, the level of community multiplexity increases by one, represented by the Kronecker delta function, which evaluates membership equivalency of the two nodes. 

Having described our multiplex methodology, which has not been previously applied to international networks of flows, we will proceed to describe the six networks and fourteen global socioeconomic indicators which we use in the core of our analysis next. 

\begin{figure}[t!]
\centering
    \includegraphics[scale=0.3]{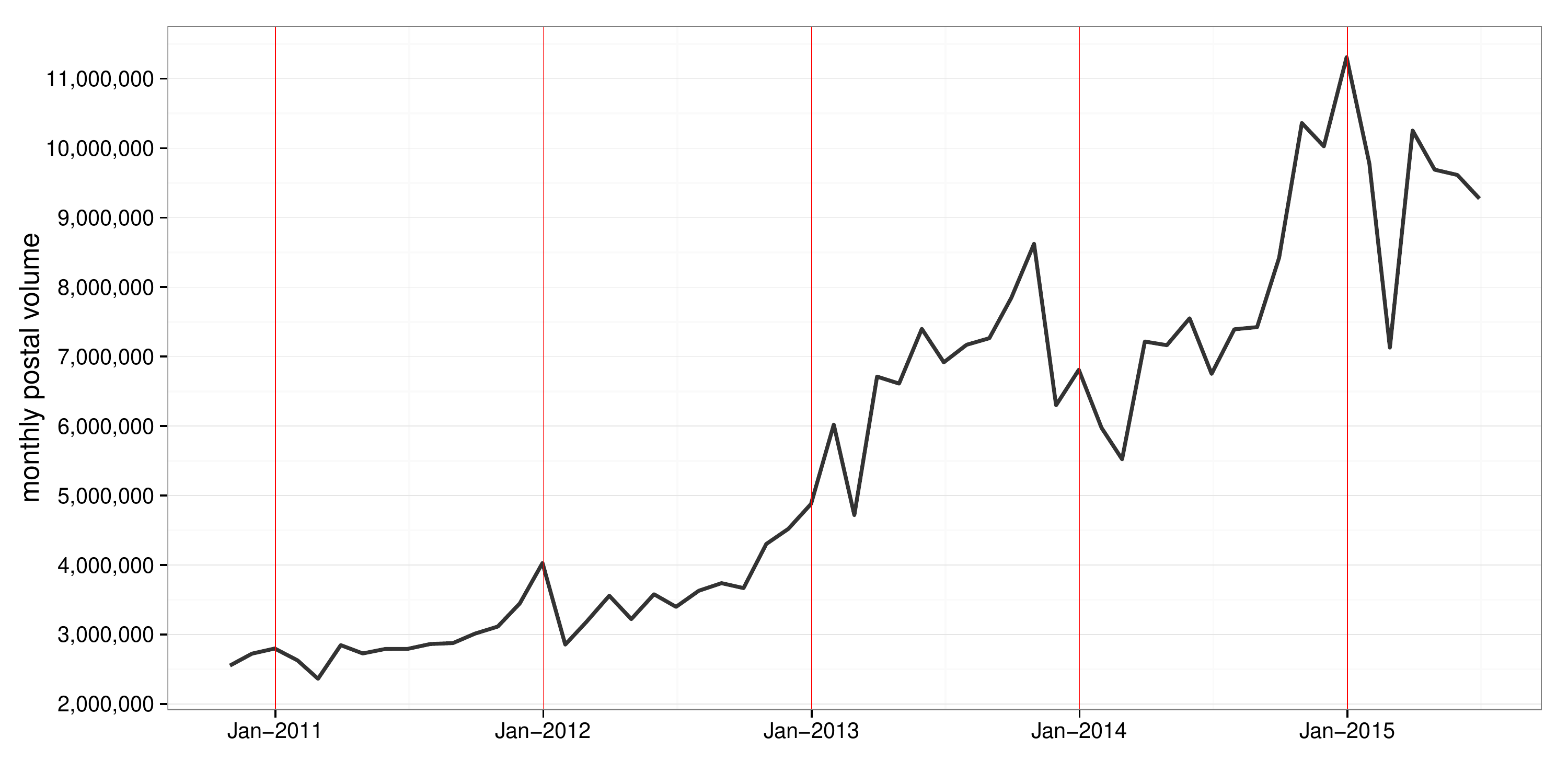} 
\caption{Global postal volume per month for the whole data period; volume is proportional to the total number of items sent between countries but does not represent its actual value due to data's sensitivity.}
\label{fig:vol}
\end{figure}

\subsection{The International Postal Network}
Although postal flows are understood to follow a distance based gravity model~\cite{anson2013}, similar to other networks describing flows, little is understood about the network properties of the postal network and how they relate to those of other global flow networks.
The International Postal Network (IPN) is constructed using electronic data records of origin and destination for individual items sent between countries collected by the Universal Postal Union (UPU) since 2010 until present. Items are recorded on a daily basis amounting to nearly 14 million records of items sent between countries. As one of the most developed communication networks on a global scale, it is a dense network with 201 countries and autonomous areas, and 23K postal connections between them, with 64\% of all possible postal connections established. The global volume of post has seasonal peaks observable in Fig.~\ref{fig:vol}. Notably, since 2010 postal activity is on the rise and this can be accounted for by the parallel growth of e-commerce~\cite{ecommerce}. This positions postal flows as a sustainable indicator of socioeconomic activity. 

\begin{figure}[t!]
\centering
     \includegraphics[scale=0.32]{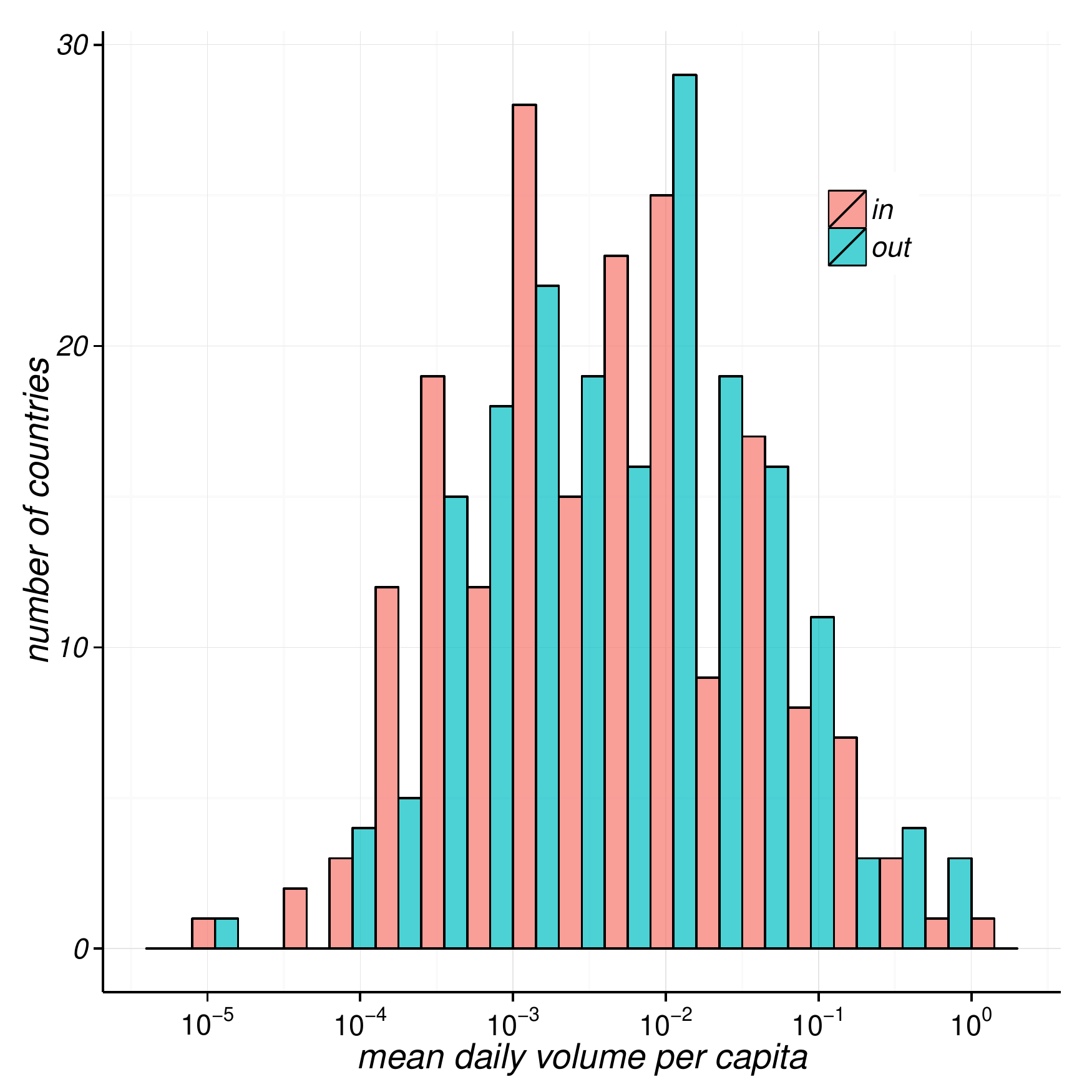} 
\caption{Average number of daily items sent (out) and received (in) per capita per country. Volume is proportional but does not represent the actual number of items exchanged due to data sensitivity.}
\label{fig:dact}
\end{figure}

In terms of daily activity, we can observe the mean relative number of daily items sent and received by countries during the period in Fig.~\ref{fig:dact}. 
This can be highly dependent on the size of the population of a country so we have normalised the volume per country's population. We use annual population statistics provided by the World Bank and collected by the United Nations Population Division. From the distribution of volume it becomes clear that the majority of countries send and receive a similar amount of post per capita, however with a number of exceptions on both ends where a few countries send and receive exceptionally low or high number of items. 

\begin{figure}[t!]
\centering
 \begin{subfigure}[b]{0.25\textwidth}
                \includegraphics[width=\textwidth]{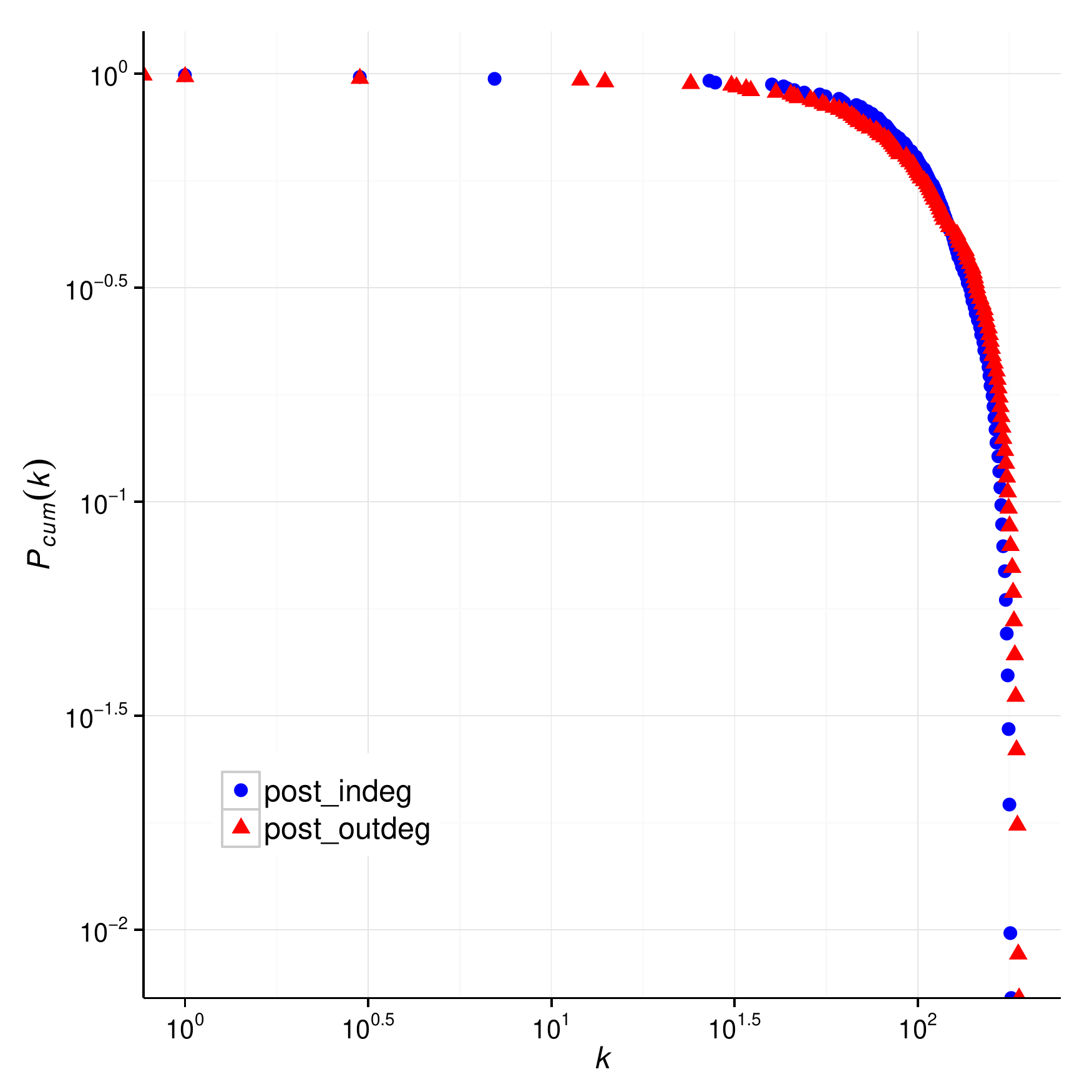}
          \caption{Unweighted degree distribution}
           \label{fig:dd}
\end{subfigure}%
           \begin{subfigure}[b]{0.25\textwidth}
                \includegraphics[width=\textwidth]{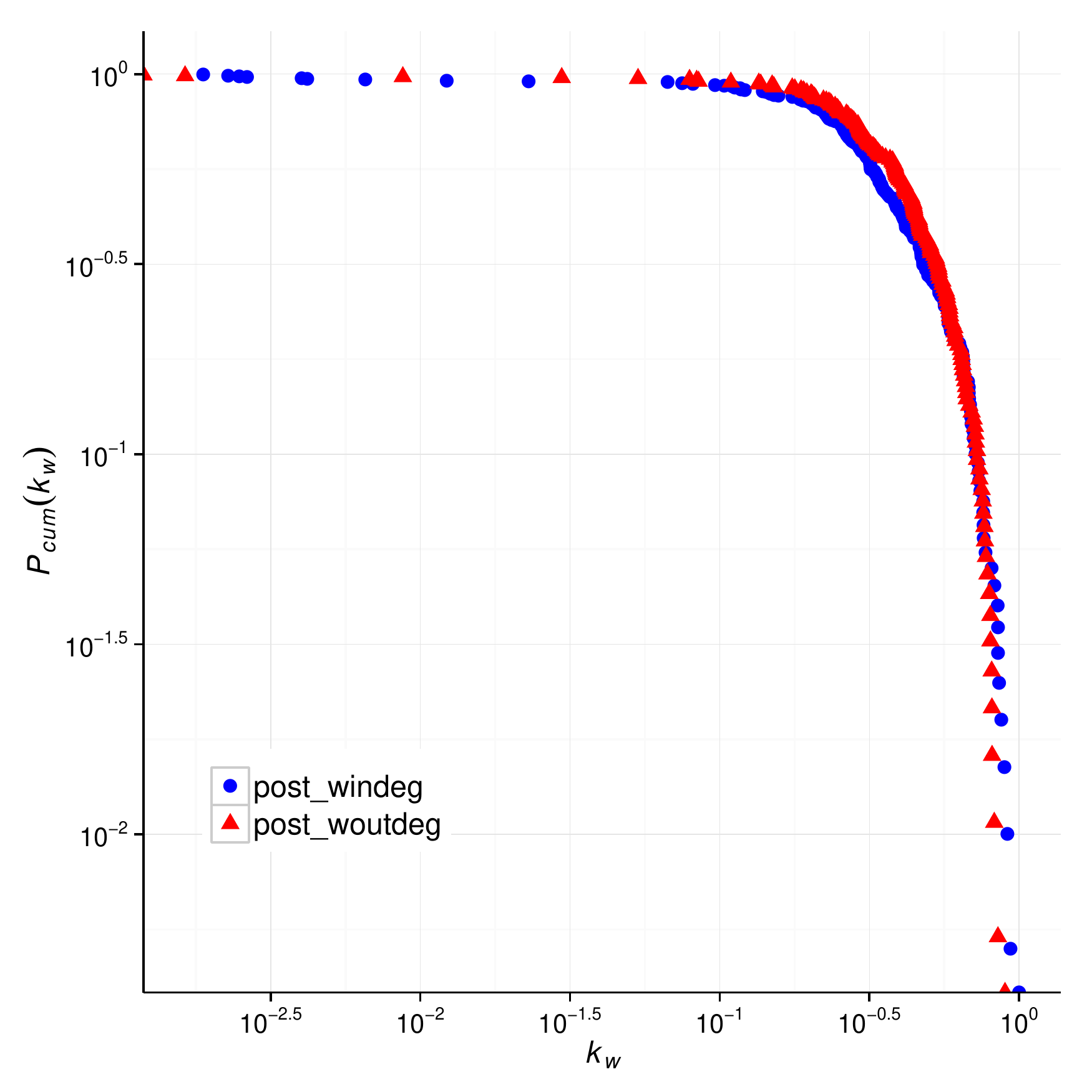}
          \caption{Weighted degree distribution}
           \label{fig:wdd}
\end{subfigure}%
\caption{International Postal Network degree distributions.}
\label{fig:degs}
\end{figure}

\begin{figure*}[t!]
\centering
\includegraphics[scale=0.52]{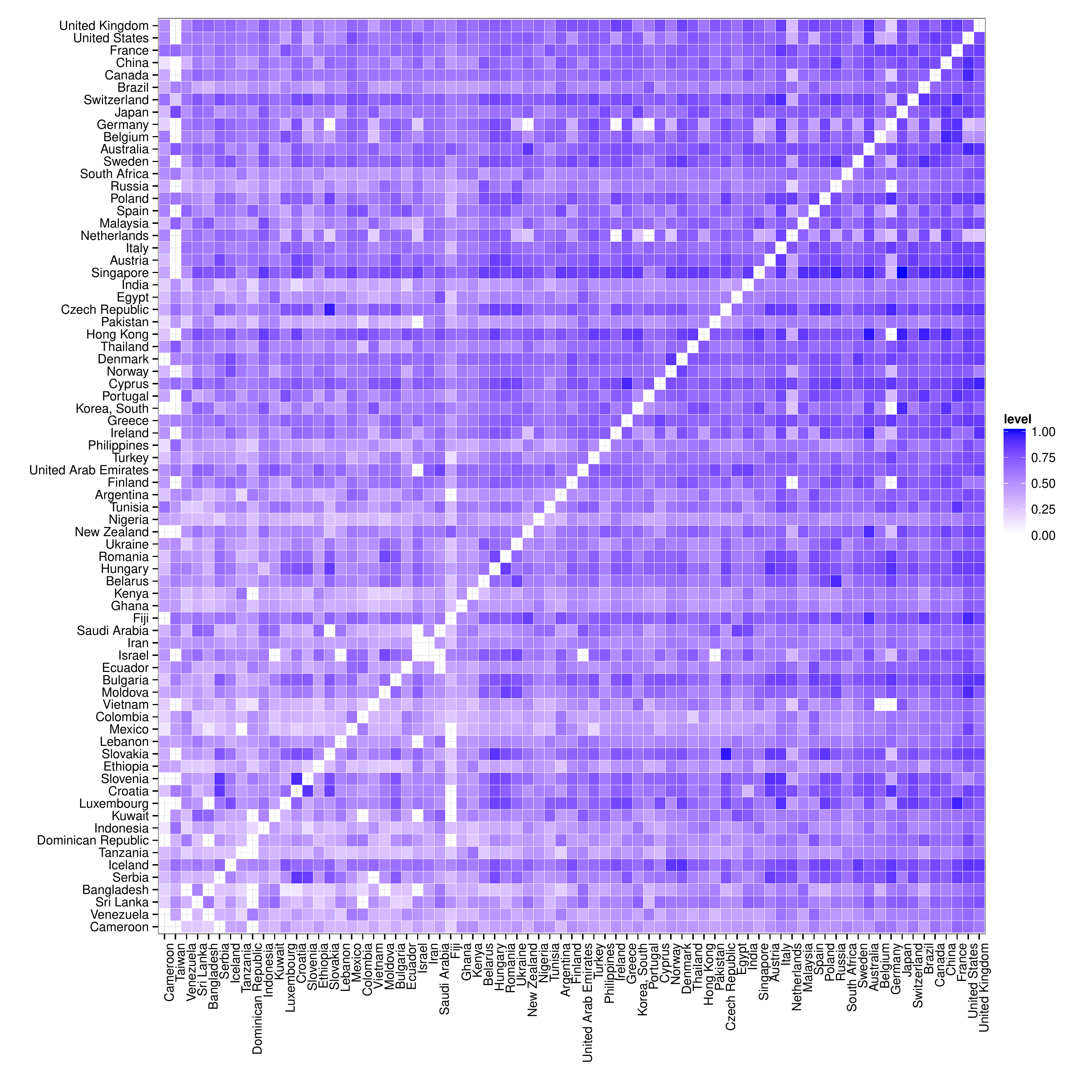}
\caption{Matrix of the intensity of connections between countries based on the number of items exchanged (higher is darker); axes are ordered by the country's unweighted postal degree (its number of postal partners); only countries with more than 120 postal partners appear for display purposes.}
\label{fig:matrix}
\end{figure*}

Next we report on the degree distributions of both the weighted and unweighted global postal graphs. The unweighted postal graph simply contains all directed edges present in the network regardless of flow volume. The weighted graph on the other hand also includes the weight of connections in the graph. We weight the network by summing the total annual volumes of directed flow between two countries, averaged over years and normalised over the population of the country of origin. We then further normalise by the maximum weight in the network, resulting in a value between 0 and 1, allowing us to compare values between networks. The weighted adjacency matrix of the top quartile of countries in terms of degree can be seen in Fig.~\ref{fig:matrix} with the US and UK having the largest numbers of postal partners. Prominent postal network countries have relatively high interaction with most of their partners, including interactions with lower ranked countries. This is related to the degree assortativity within the postal network, discussed in the following section. Further, both weighted and unweighted degree distributions are shown in Fig.~\ref{fig:degs}, as the complementary cumulative probability function (CCDF). We can see in Fig.~\ref{fig:degs}B that the in and out degrees are relatively balanced in both instances and that about 50\% of countries have more than 100 postal partners. The weighted degree in Fig.~\ref{fig:degs}A follows a similar pattern, which means that countries tend to interact equally proportional to the number of  their postal partners. In the following section, we will compare the postal network properties to other flow networks. 

\subsection{Other global flow networks}

This work builds upon previous efforts using global flow networks to present novel data sources for international development efforts such as the IPN and to demonstrate a holistic view of several distinct flow networks. We consider five networks, which have been previously studied independently, along with the IPN. We will now describe these networks and compare their network properties in the following section.

\paragraph{The World Trade Network} The trade network is constructed from records maintained by the UN Statistics Division in the Comtrade Database and provided by the Atlas Project~\footnote{https://atlas.media.mit.edu/about/data/sources/} and contains the number and value of products traded between countries classified by commodity class.

\begin{figure*}[t!]
\centering
 \begin{subfigure}[b]{0.25\textwidth}
                \includegraphics[width=\textwidth]{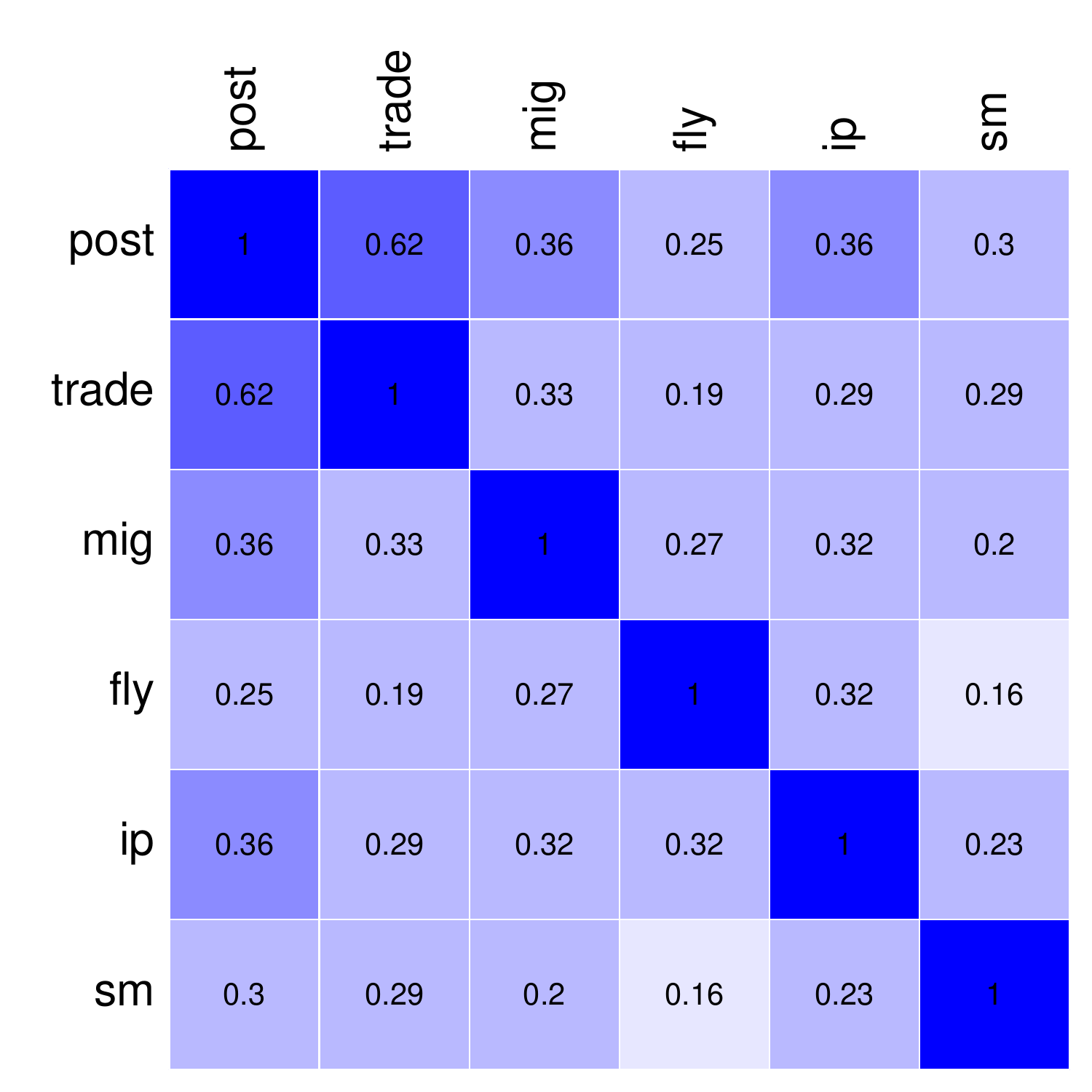}
          \caption{Jaccard edge overlap coefficient}
           \label{fig:comp1}
\end{subfigure}%
           \begin{subfigure}[b]{0.25\textwidth}
                \includegraphics[width=\textwidth]{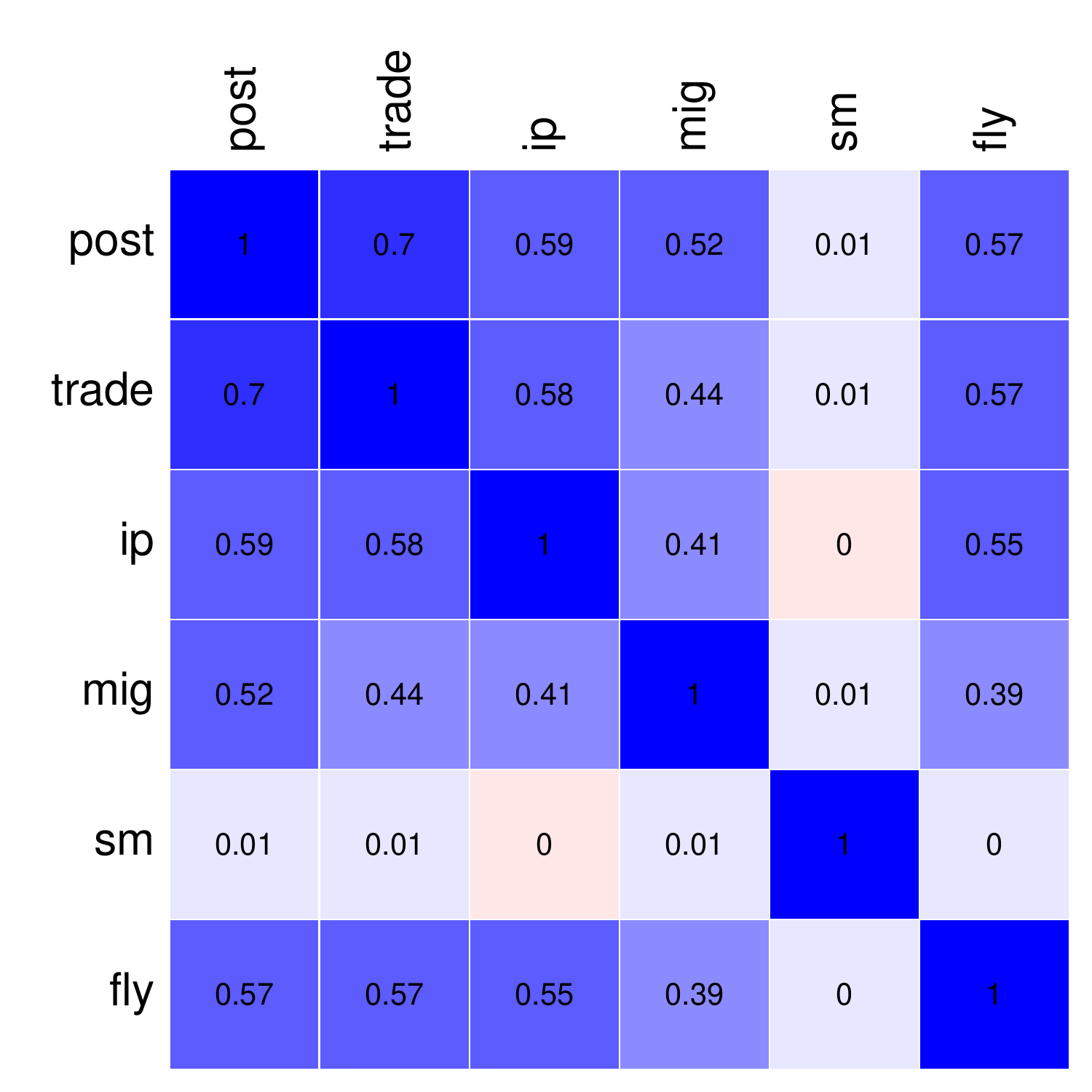}
          \caption{Weighted  edge correlations}
           \label{fig:comp2}
\end{subfigure}%
           \begin{subfigure}[b]{0.25\textwidth}
                \includegraphics[width=\textwidth]{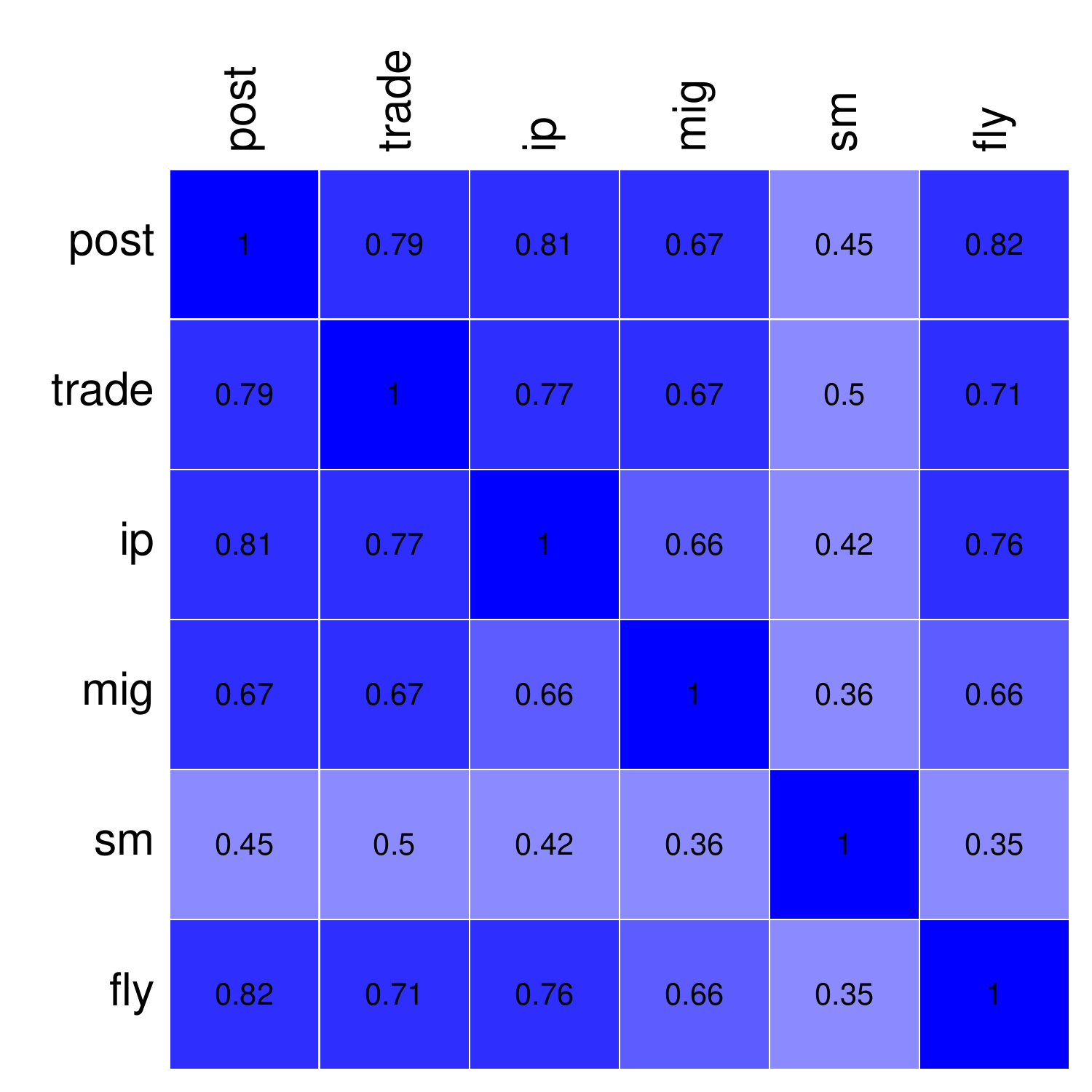}
          \caption{Unweighted in degree correlations}
           \label{fig:comp3}
\end{subfigure}%
           \begin{subfigure}[b]{0.25\textwidth}
                \includegraphics[width=\textwidth]{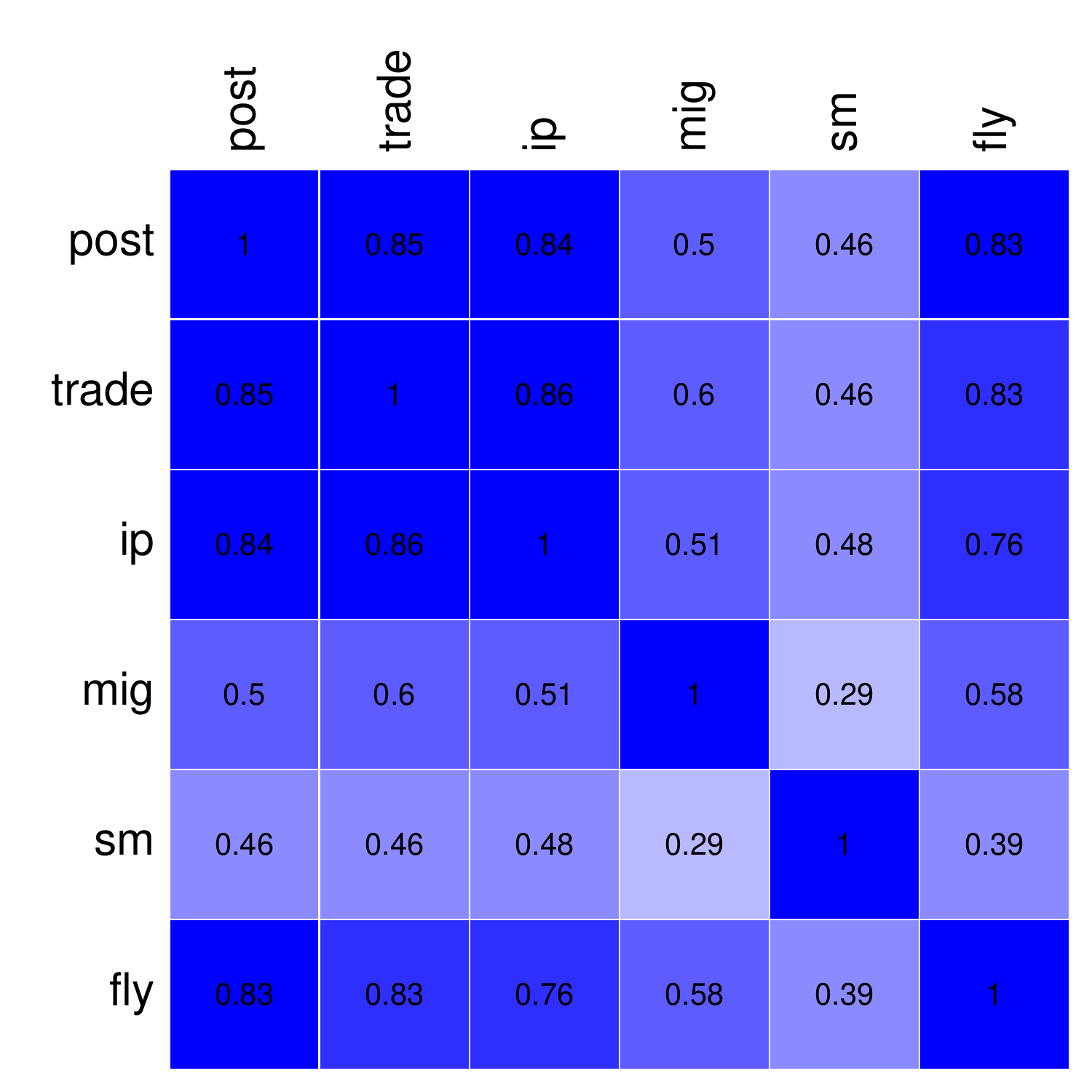}
          \caption{Unweighted out degree correlations}
           \label{fig:comp4}
\end{subfigure}%
\caption{Comparative analysis of the six network in terms of Jaccard overlap, percent shared edges, edge weight correlation and in and out degree correlations calculated using the Spearman ranked correlation method. All results are statistically significant (p<0.05).}
\label{fig:comp}
\end{figure*}

\paragraph{The Global Migration Network}

This is compiled from bilateral flows between 196 countries as estimated from sequential stock tables. It captures the number of people who changed their country of residence over a five-year period. This reflects \textit{migration transitions} and not short term movements. This data is provided by the global migration project.~\footnote{http://www.global-migration.info/} 

\paragraph{The International Flights Network} The flights data is collected by 191 national civil aviation administrations and compiled by the International Civil Aviation Organisation (ICAO)\footnote{http://www.icao.int/Pages/default.aspx}. These tables detail, for all commercial passenger and freight flights, country of origin and destination and the number of flights between them. \cite{airline}.

\paragraph{The IP Traceroute Network} This city to city geocoded dataset is built from traceroutes in the form of directed IP to IP edges collected in a crowdsourced fashion by volunteers through the DIMES Project.~\footnote{http://www.netdimes.org/} The project relies on data from volunteers who have installed the measurement software which collects origin, destination and number of IP level edges which were discovered daily. We aggregate this data on a country to country basis and use it to construct an undirected Internet topology network, weighted by the number of IPs discovered and normalised by population as all other networks. The data collection methods are described in detail in the founding paper of the project~\cite{dimes2005}. The global mapping of the Internet topology provides insight into international relationships from the perspective of the digital infrastructure layer.

\paragraph{The Social Media Density Network} is constructed from aggregated digital communication data from the Mesh of Civilizations project, where Twitter and Yahoo email data is combined to produce an openly available density measure of the strength of digital communication between nations~\cite{bogdan2015}. This measure is normalised by the population of Internet users in each country and thus is well aligned with the rest of the networks we use. It also blends data from two distinct sources and thus provides greater independence from service bias. Because the study considers tie strength, it only includes bi-directed edges in the two platforms where there has been a reciprocal exchange of information and therefore this network is undirected.\\

In the following analysis we compare these networks and use multiplexity theory to extract knowledge about the strength of connectivity across them. We will distinguish between single layer and multiplex measures, which will allow us to observe to a deeper extent the international relationships and the potential for using global flow networks to estimate the wellbeing of countries in terms of a number of socioeconomic indicators (summarised in Table~\ref{tab:indi}).

\begin{table} [t!]
\tiny
\centering
\begin{tabular}{|l|l|l|l|}
\hline
\textbf{Abbreviated} & \textbf{Full name} & \textbf{Description} & \textbf{Source}\\ \hline
\hline
\multirow{3}{*}{GDP} & & Aggregate measure  of & \\
& Gross Domestic Product & production on a& The World Bank\\
& & on a per capita basis & \\
\hline
\multirow{3}{*}{LifeExp} & Life Expectancy & Life expectancy since & \\
& & birth in years & The World Bank\\
\hline
\multirow{3}{*}{CPI} &  & Perceived levels of corruption,& \\
& Corruption Perception Index &  as determined by expert& Transparency International\\
& & assessments and opinion surveys &\\
\hline
\multirow{3}{*}{Happiness} &  &Survey of the state   &  \\
& Happiness Score & of global happiness perceptions & Gallup World Poll\\
& & &\\
\hline
\multirow{3}{*}{Gini.Idx} &  &  Income & \\
& Gini Index & inequality on a& The World Bank\\ 
& & national level  &\\
\hline
\multirow{3}{*}{ECI} &  & Holistic measure of  &The Observatory  \\
& Economic Complexity Index & the production characteristics & of Economic Complexity\\
& & of large economic systems&\\
\hline
\multirow{3}{*}{LitRate} & & Percent of adult population & \\
& Adult Literacy Rate & who are literate & UNESCO\\
&&&\\
\hline
\multirow{3}{*}{PovRate} & & Percent of population & \\
& Poverty Rate & living bellow national poverty & The World Bank\\
& & threshold &\\
\hline
\multirow{3}{*}{EdRate} &  & Percent of population &  \\
& Education Rate & who have completed & The World Bank\\
& & primary school &\\
\hline
\multirow{3}{*}{CO2} &  & Carbon dioxide &  Carbon Dioxide Information \\
& Emissions of carbon dioxide & in billions of metric tonnes & Analysis Center\\
&&per capita&\\
\hline
\multirow{3}{*}{FxPhone} & & Percent of population & \\
& Fixed Phone Rate & living in households with & Int Telecommunication Union \\
& & a fixed phone line & \\
\hline
\multirow{3}{*}{Inet} &  & Percent of population & \\
& Internet penetration & who have accessed & Int Telecommunication Union\\
&&the Internet in the past 12 months &\\
\hline
\multirow{3}{*}{Mobile} & & Percent of population & \\
& Mobile cellular subscriptions & who have a mobile cellular& Int Telecommunication Union\\
& & subscription &\\
\hline
\multirow{3}{*}{HDI} & & Composite statistic of life expectancy, &\\
& Human Development Index & education, and income & UNDP\\
&&per capita indicators&  \\
\hline
\end{tabular}
\caption{Description and source of the fourteen indicators we try to approximate using flow network measures}
\label{tab:indi}
\end{table}

\section{Results}

In order to understand the multiplex relationships of countries through flows of information and goods in context, we first compare all flow networks together. We then present their respective and collective ability to approximate crucial socioeconomic indicators and finally perform a network community analysis of individual networks and their multiplex communities where the most socioeconomically similar countries can be found. 

\begin{table*} [t!]
\centering
\begin{tabular}{l*{8}{l}c}
\hline
\textbf{network} & \textbf{weight} & \textbf{years} & $|\mathbf{V}|$ & $|\mathbf{E}|$ & $<\mathbf{k}>$  & \textbf{assort} & $\mathbf{d}$ & \textbf{cc}\\
\hline
Post & postal items & 2010 -- 15 & 201 & 22,280 & 110.85 & -0.26 & 0.55 & 0.79 \\
\hline
Trade & export value & 2007 -- 12 & 228 &  30,235 & 132.6 & -0.39 & 0.58 & 0.84\\
\hline
Migration & migrants & 2005 -- 10 & 193 &  11,431 & 59.22 & -0.33 & 0.31 & 0.68\\
\hline
Flights & flights & 2010 -- 15 & 223 & 6,425 & 28.81 & -0.1 & 0.13 & 0.49\\
\hline
IP  & IPs & 2007 -- 11 & 225 & 9,717 & 43.19 & -0.42 & 0.19 & 0.6\\
\hline
SM & density & 2009 & 147 & 10,667 & 145.13 & -0.02 & 0.98 & 0.99\\
\end{tabular}
\caption{Network Properties: number of nodes, number of edges, average (out) degree, degree assortativity, network density, average clustering coefficient.}
\label{tab:nets}
\end{table*}

\subsection{Comparing networks}

Although each of the five networks previously described apart from the International Postal Network (IPN) has been studied separately, there has not been a comparative analysis of all. In Table~\ref{tab:nets}, we list the network properties of all six network separately. The number of nodes or countries exceeds 195(6) due to differing lists of member states providing statistics to each authority. In terms of weights, although distinct for each network, it is also a value of volume that is flowing between areas. While there are small discrepancies between the years of each network, most networks cover a five year period, with the exception of the Social Media network which is from a single year. 
The volume of interaction between two countries is therefore averaged over the number of years for each network.

We weight all networks by normalising the raw volume of interaction described above by the population of each respective country of origin and rescaling all weights across networks within the same range [0,1] by dividing by the maximal weight, as we did for the postal network in the previous section. We compute the out degree for each network in a standard way as for the postal network, as well as the degree assortativity (Pearson correlation between the degrees of two connected countries), the network density and clustering coefficient. The assortativity coefficient determines to what extent nodes in the network have mixing patterns that are determined by their degree. Positive assortativity means that nodes with high degree tend to connect to other nodes with high degree, whereas a negative assortativity means that nodes with high degree tend to connect with others with lower degree, which is the case for most of the six networks.

Although all networks differ in size and average degree, they have relatively high clustering coefficients, reflecting a general tendency for countries to cluster together in global networks. This clustering however is not based on the importance of a node (its degree) since the assortativity coefficients for all networks are low or negative, suggesting that global networks are dissassortative and therefore higher degree nodes tend to connect to lower degree nodes.

Fig.~\ref{fig:comp} presents a comparative analysis between the six networks. We refer to them for short as: post, trade, ip, mig, sm and fly.  We use the Jaccard coefficient to compute the overlap of edges in Fig.\ref{fig:comp}A, where we have the number of edges that exist on both networks over the union of edges on the two networks. The highest Jaccard overlap is between the postal and trade networks, the two densest networks. The rest of the networks however are not strongly overlapping in terms of edges, which implies that each distinct network layer provides a non-trivial and complementary view of how countries connect.
The correlation between edges in Fig.~\ref{fig:comp}B reveals that the volume of flow of goods, people, and information is correlated for those edges which exist on both networks. A notable exception is the digital communications network (sm), which is entirely uncorrelated in terms of density with any other network. This means that countries likely connect in unexpected ways on social media and email.

When considering the degree of a country as an indicator of its position in the network, we find that there are high correlations between the in and out positions of countries in Fig.~\ref{fig:comp}C and Fig.~\ref{fig:comp}D. Although lower, the social media network is also correlated with the others. We should  note that this is likely due to the smaller overlap between edges but for the nodes present across networks, we find that there is a strong correspondence between their positions in the different networks. Next we will explore how well different degree metrics approximate the socioeconomic indicators described above.

\subsection{Approximating indicators}

Timely statistics on key metrics of socio-economic status are essential for provision of services to societies, in particular marginalised populations. The motivation for this measurement varies from social resilience in the event of natural or man-made disasters to ensuring social rights such as education and access to information. While national governments typically administer their territories and allocate resources in terms of sub national divisions, international organisations such as the United Nations and the World Bank, as well as regional organisations and blocs such as the Economic Council or Latin American and the Caribbean and the African Union invariably partition populations under nation states. In this context, the nation state is the primary geographical entity considered for  funding, planning and allocation of resources for development. Despite the importance of accurate statistics to quantify the state of a country and progress towards favourable socio-economic outcomes, regular and reliable measurement is difficult and costly particularly in low income countries.

\begin{figure}[t!]
\centering
\includegraphics[scale=0.4]{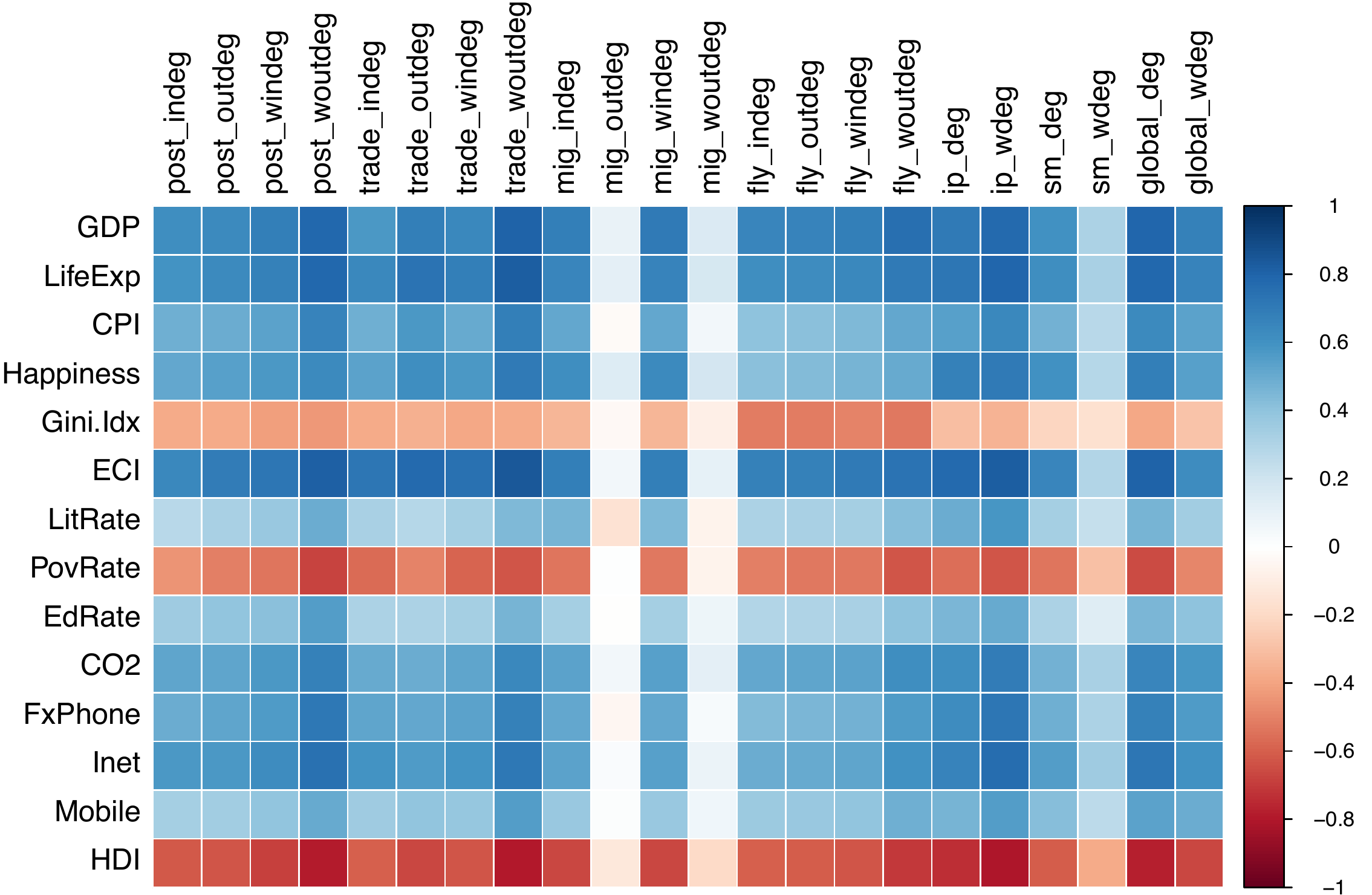}
\caption{Spearman rank correlations between global flow network degrees and socioeconomic indicators}
\label{fig:cormatrix}
\end{figure}

With this in mind, in this section we compare the positions of countries within the different networks discussed previously to the values of several socioeconomic indicators. Fig.~\ref{fig:cormatrix} shows the Spearman rank correlation between the network degrees of the six networks (in and out degree, and weighted in and out degree) and various socio-economic indicators: GDP, Life expectancy, Corruption Perception Index (CPI), Internet penetration rate, Happiness index, Gini index, Economic Complexity Index (ECI), Literacy, Poverty, $CO_2$ emissions, Fixed phone line penetration, Mobile phone users, and the Human Development Index. These indicators and their significance for the international development agenda are described in detail in the data section (see Table~\ref{tab:indi}). 

For each of the six networks, we compute the network degree, defined as the sum of the neighbours for both incoming and outgoing connections where directed. This reflects how well connected a country is in a particular network. We also take into account the amount of connectivity by computing the weighted incoming and outgoing degrees on each network, defined as the sum of the normalised flows from all neighbours and reflecting the volume of incoming and outgoing flows. In addition to these standard single-layer network metrics, we define and compute the \emph{global degree} of a country, which takes into account connectivity across all networks. 

All degrees of single networks and the global degree appear vertically in Fig.~\ref{fig:cormatrix} and all indicators appear horizontally. In general, weighted outgoing degrees on the single networks perform best for the postal, trade, ip and flight networks. An exception from the physical flow networks is the migration network, where the incoming migration degree is more correlated with the various indicators. The best-performing degree across is the global degree. This suggests that looking at how well connected a country is in the global multiplex can be more indicative of its socioeconomic profile than looking at single networks.

\begin{figure*}[t!]
\centering
        \begin{subfigure}[b]{0.3\textwidth}
                \includegraphics[width=\textwidth]{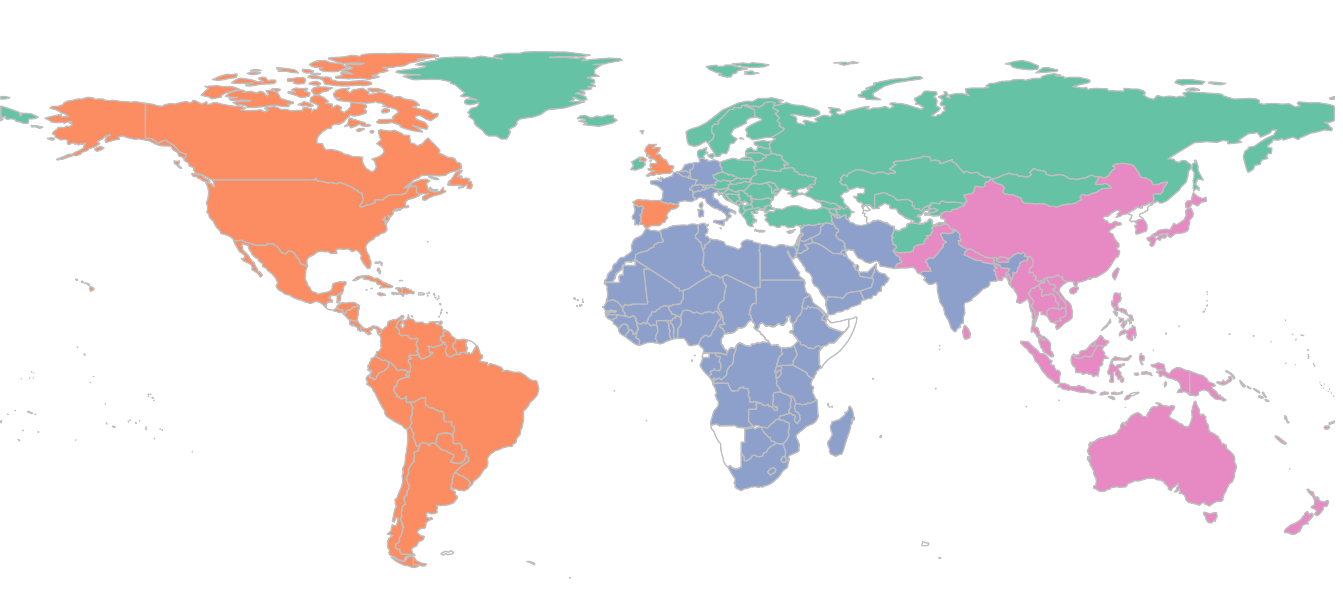}
                \caption{Postal network}
                \label{fig:ideg}
        \end{subfigure}
        \begin{subfigure}[b]{0.3\textwidth}
                \includegraphics[width=\textwidth]{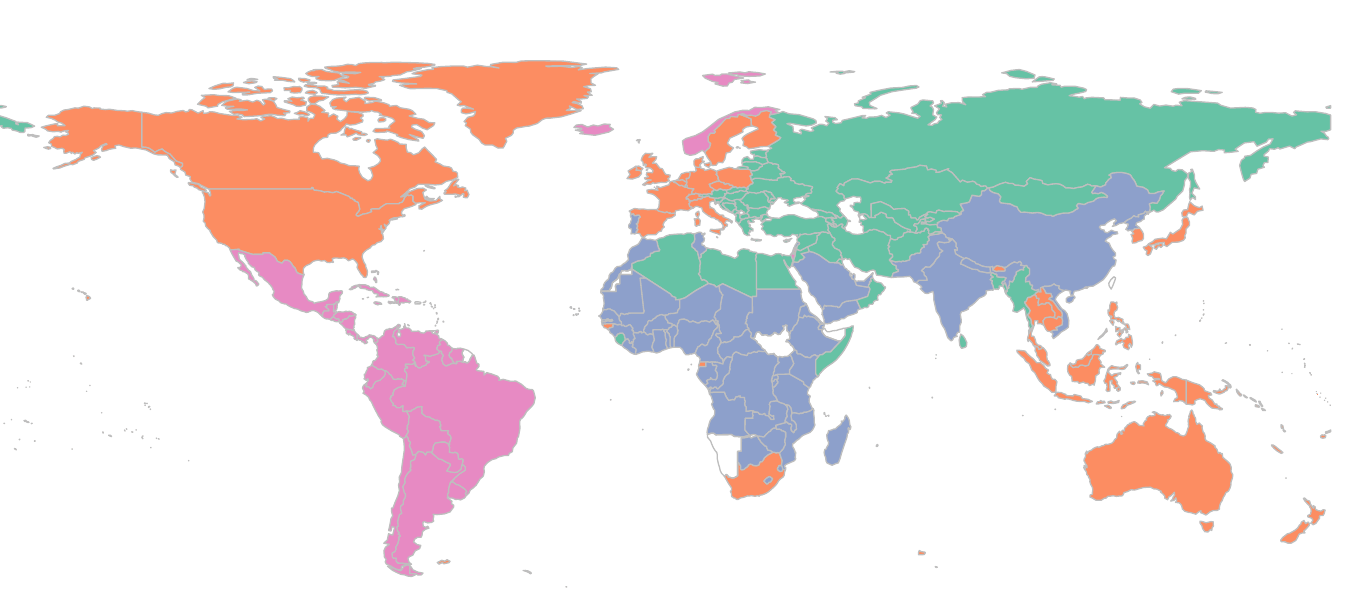}
                \caption{Trade network}
                \label{fig:udeg}
        \end{subfigure}%
                \begin{subfigure}[b]{0.3\textwidth}
                \includegraphics[width=\textwidth]{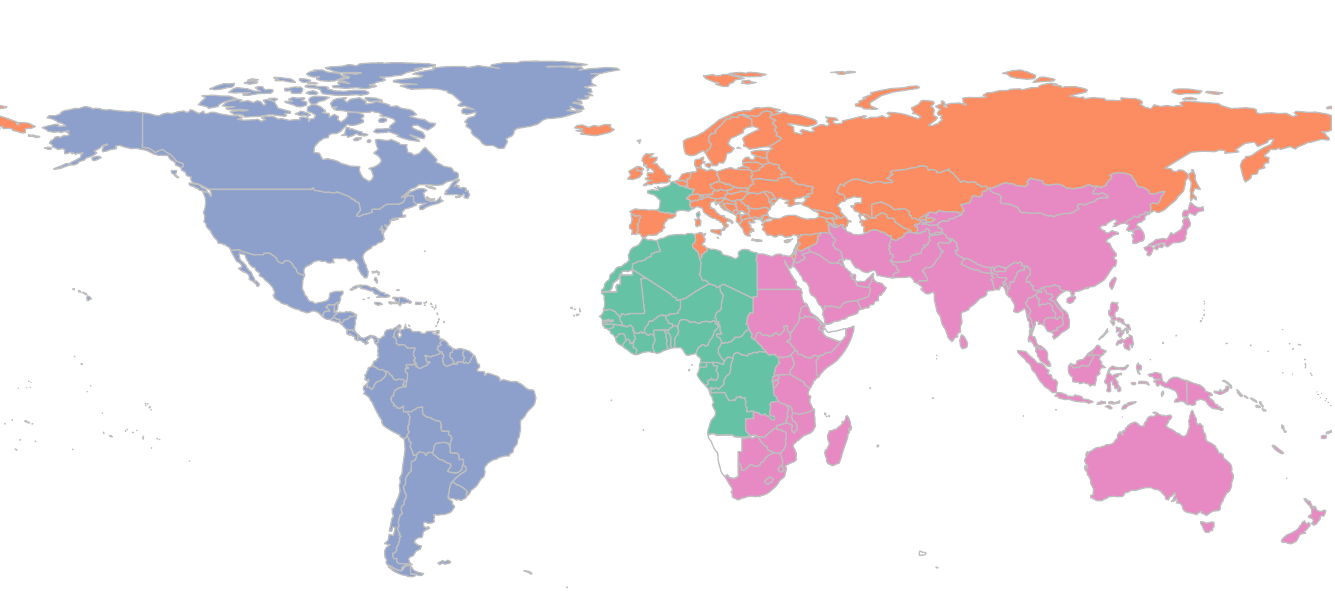}
                \caption{Flight network}
                \label{fig:uvi}
        \end{subfigure}
             
       \begin{subfigure}[b]{0.3\textwidth}
                \includegraphics[width=\textwidth]{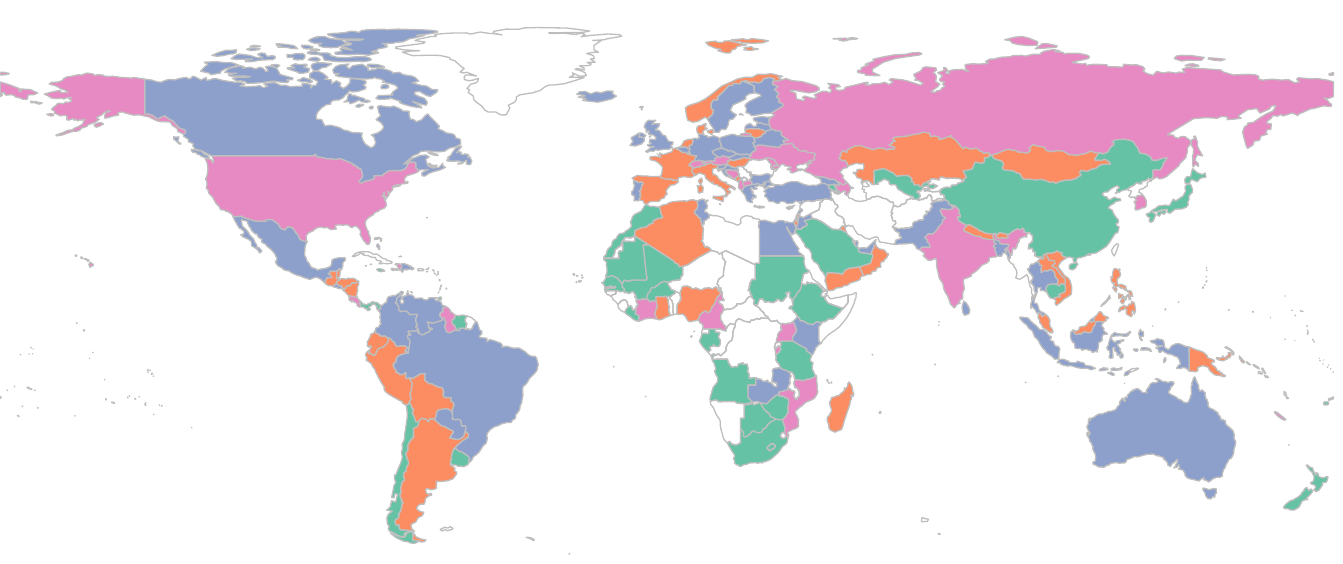}
                \caption{Digital Communications network}
                \label{fig:sm}
        \end{subfigure}%
                      \begin{subfigure}[b]{0.3\textwidth}
                \includegraphics[width=\textwidth]{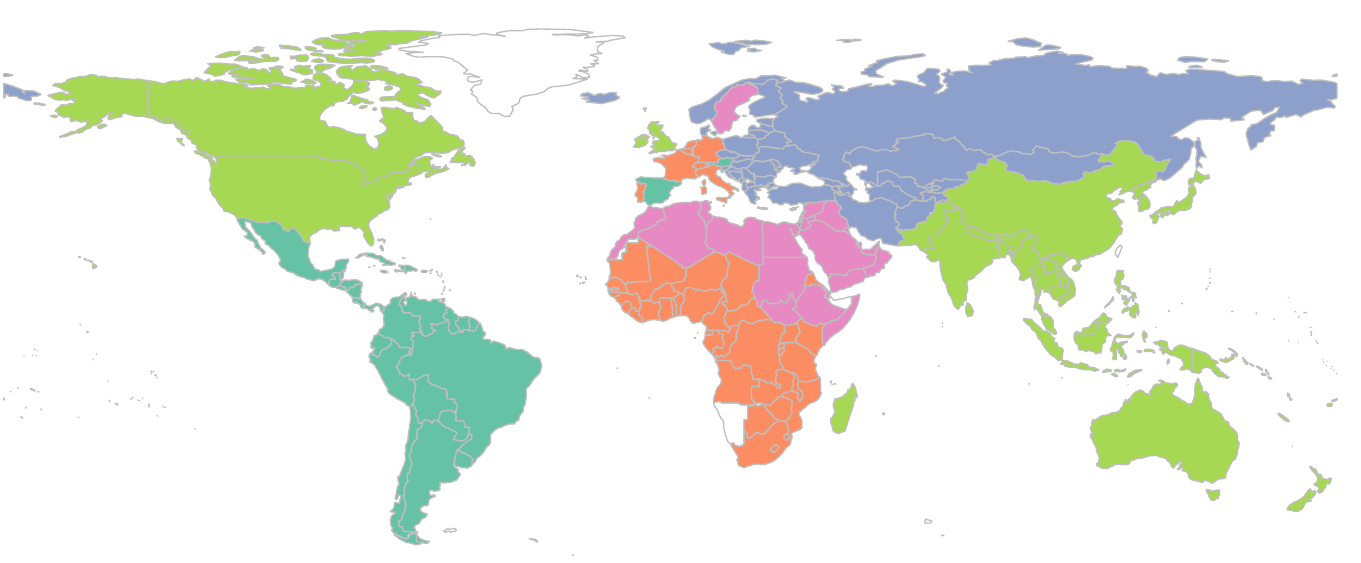}
                \caption{Migration network}
                \label{fig:sm}
        \end{subfigure}%
         \begin{subfigure}[b]{0.3\textwidth}
             \centering
                \includegraphics[width=\textwidth]{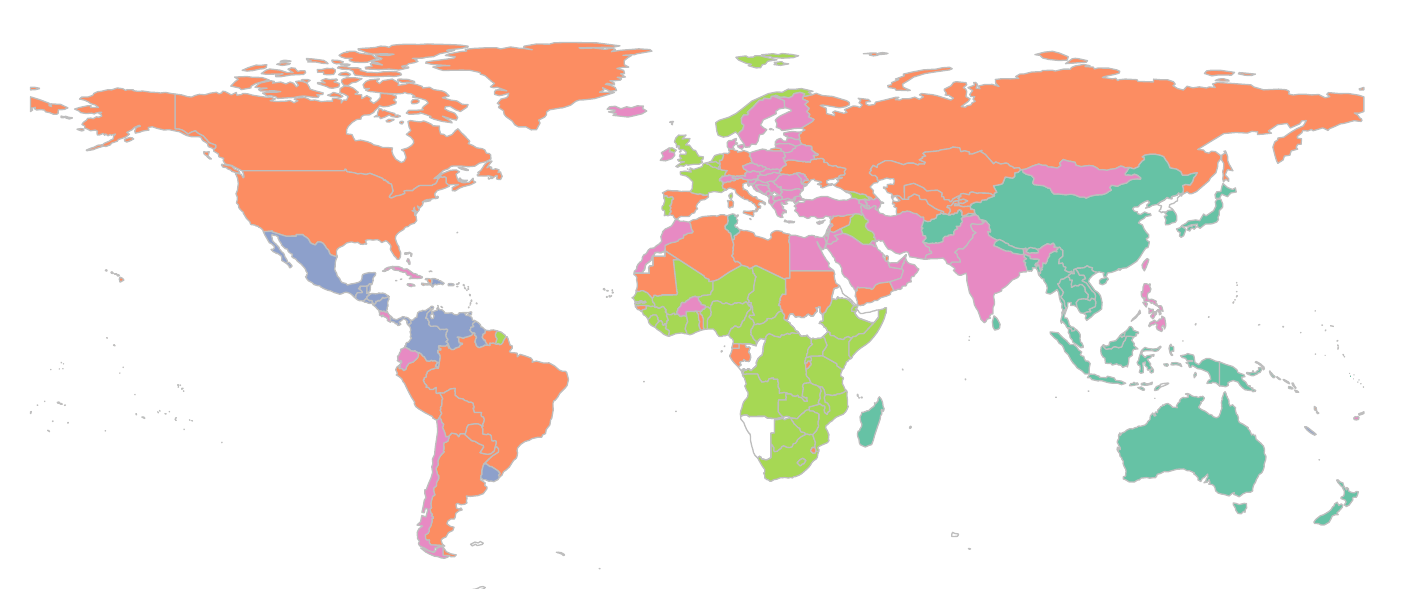}
                \caption{IP network}
                \label{fig:odeg}
        \end{subfigure}%
         	\centering
        \caption{Country community membership for each network.}
        \label{fig:coms}
\end{figure*}

The GDP per capita and life expectancy are most closely correlated with the global degree, closely followed by the postal, trade and ip weighed degrees. This shows a relationship between national wealth and the flow of goods and information. The perception of corruption index (CPI) however, is most positively correlated with the out weighted degrees of the postal and trade networks, followed by the IP network but not so strongly with their out degrees, similar to their relationship with the happiness index. This signifies that less corrupt and more happy countries have greater outflows in those respects. On the other hand, the Gini Index of inequality is distinctly most negatively correlated with the flight network, which means that countries with greater inequality have less incoming and outgoing flight connections. The ECI index is equally highly correlated with most network degrees, and especially the global degree, trade, ip and post degrees. Literacy, Education and mobile phone users per capita were more weakly correlated across than other indicators, which means that there may be better predictor variables beyond the scope of this work for those indicators. Fixed phone line households, Internet penetration and CO2 emissions, however, are positively correlated with the global degree, followed by the postal and ip degrees. This indicates the importance of global connectivity across networks with respect to these factors.

Similarly to GDP, the rate of poverty of a country is best represented by the global degree, followed by the postal degree. The negative correlation indicates that the more impoverished a country is, the less well connected it is to the rest of the world. Finally, one of the most strongly correlated indicators with the various degrees is the Human Development Index (HDI), low human development (high rank) is most highly negatively  correlated with the global degree, followed by the postal, trade and ip degrees.  This shows that high human development (low rank) is associated with high global connectivity and activity in terms of incoming and outgoing flows of information and goods.
One notable observation is that the ip, postal and trade weighted out network degrees all have similar correlation patterns with the various indicators, the commonality between these networks is that they express the flow of resources from a country. Another observation is that weighted social media and migration outflow are weak predictors of the explored indicators. Because most indicators are related to each other, e.g., high GDP indicates low Poverty or high HDI indicates Happiness, when a degree is a predictor of one, it tends to be a good predictor of the others. 

In this section we have shown that network science can provide reliable and easy to compute approximations of various indices and that connectivity between countries determines their position in global flow networks which relate to the success of their socioeconomic properties. Next, we will look at the community structure of countries across networks and evaluate their community multiplexity to show that countries with similar socioeconomic profiles tend to cluster together, much like in social networks.

\subsection{Global Community Analysis}

In the previous section we related network measures to various socioeconomic indicators, showing that metrics such as the network degree can be used to estimate wellbeing at a national level. In this section, we further examine the relationship between countries and the way in which they cluster into communities across networks and the relationship of those communities to the various socioeconomic indicators. We use the Louvain modularity optimisation method~\cite{louvain} for community detection in each individual network, which takes into account the tie strength of relationships between countries and finds the optimal split in terms of disconnectedness in the international network. This returns between 4-6 communities for each network, the geographical distribution of which is shown in Fig.~\ref{fig:coms}.

\begin{figure*}[t!]
\centering
\includegraphics[scale=0.5]{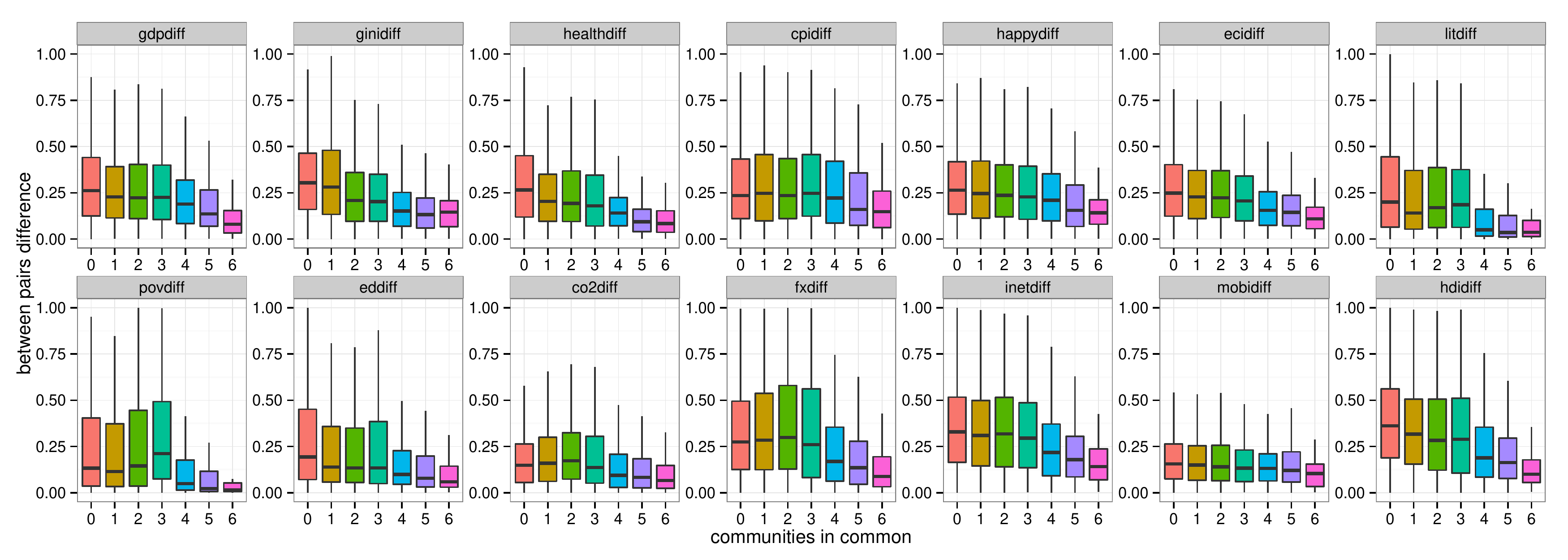}
\caption{Socioeconomic difference margin between countries who share communities in the global flow networks.}
\label{fig:diff}
\end{figure*}

Although communities naturally seem to be very driven by geography in physical flow networks, this is not the case in digital networks where communities are geographically dispersed. This is an indication of the difference in the way countries connect through post, trade, migration and flights rather than on the IP and social media networks.
However, \emph{what does it mean for two countries to be both members of the same network community?} Common community membership indicates a level of connectedness between two countries, which is beyond the randomly expected for the network. It is often observed that nodes in the same communities share many similar properties, therefore it can be expected that \emph{pairs of nodes which share multiple communities across networks} are even more similar. In this work, we measure the overlap in pairwise membership between pairs of countries across our six networks. 

Our hypothesis is that countries that are paired together in communities across more networks are more likely to be socioeconomically similar. We measure similarity here as the absolute difference between each indicator from the previous section for two countries and plot that against their community multiplexity. For example, the United States has an average life expectancy of 70 years, whereas Afghanistan has an average life expectancy of 50, the absolute difference between the two is 20 which represents low similarity when compared to the United Kingdom's life expectancy of 72 for this indicator.
In Fig.~\ref{fig:diff}, we can observe the variations in similarity for countries with different levels of community multiplexity. What is immediately striking is that countries that share a maximal number of communities and therefore exhibit the greatest community multiplexity, have the smallest margin of difference across all indicators. This suggests that \emph{countries with the highest community multiplexity have a very similar socioeconomic profile.} This is confirmed by a two-sample Kolmogorov-Smirnov test between the distributions of differences in each indicator for pairs sharing different numbers of communities. Although the KS statistic is lower between groups sharing 0 and 1 communities (apx. 0.1 for all indicators and p-value \textless 0.01), it is very high for groups between 1 and 6 communities (0.4 and above,  p-value \textless 0.01), except for mobile phone penetration.

Further to this observation, in most indicators there is a very strong significance in the level of community multiplexity - \emph{the higher the community multiplexity between two countries, the smaller the difference between their socioeconomic profiles.} There are notable exceptions to this such as the mobile phone penetration ratio, where it appears that beyond the highest level of multiplexity, all other countries are relatively similar in this aspect with low variation even for those pairs of countries which share no communities. For all other indicators such as GDP, Literacy ratio, HDI and Internet penetration, there is a dramatic increase in similarity past a community multiplexity of 3. Ultimately, these similarities can be used to estimate the wellbeing of countries for which it is unknown but can be estimated from its neighbours.

\section{Discussion}

Big data is often related to real-time data captured through the Internet or social networks. However, the digital divide makes access to big data insights for development more challenging in the least developed and many developing and emerging countries. Can we rely on other networks to overcome these critical data gaps in view of better measuring and monitoring developmental progress? This is particularly important following the United Nations adoption of the Sustainable Development Goals (SDG) in September 2015, made of 17 goals, 169 targets and almost 200 universal indicators, each of them calling for regular and increasingly disaggregated monitoring in every country during the 2016-30 period. This commitment invites a nuanced discussion on the nature and importance of measurement, inference and triangulation of data sources. This discussion is particularly prescient in the face of complex intertwined developmental challenges in an age of increased globalisation, economic interdependence and climate change.

The work presented above has clearly shown the value of measuring, comparing, and combining metrics of global connectivity across six different global networks in order to approximate socioeconomic indicators and to identify network communities with similar connectivity profiles.  We have shown how both global digital and physical network flows can contribute to support a better monitoring of SDG indicators, as illustrated by the high correlation between Internet and postal flows on the one hand, with an exhaustive list of socioeconomic indicators on the other hand.

We also note the considerable potential, exposed here, for future applications of postal flow data. While we have here restricted our analysis to country-level relations, postal flows allow for socio-economic mapping on a sub-national level which can inform development programmes on a practical level. An additional dimension to be explored - that is beyond the scope of this paper is temporal analysis which, combined with the multiplex network model presented above, could provide early warning of economic shocks and their propagation~\cite{shocks}.

Interestingly, despite the ease of \textit{digital} interactions and subsequent evidence that `distance is dead' ~\cite{distancedead}, \textit{physical} networks, particularly the global postal, flight and migration networks, are still stronger candidates for proxy variables in case of missing data than digital networks such as the Internet or social media. These networks not only reach populations excluded from access to digital communications, but are also associated with the highest number of country pairs sharing relatively similar socioeconomic patterns, in turn opening numerous ways of completing missing data with proxy variables.  In the digital era, greater granularity and frequency of analysis and monitoring of SDGs can, paradoxically, be achieved through global physical networks data. We expect that the value as proxies for the digital communication networks will increase as they mature, expand and become more accessible. In the near future, both physical and digital networks will need to be combined to optimise monitoring efforts. In that sense, the emergence of the Internet of things (IoT) could play a critical role by making even more fuzzy the frontiers between the digital and physical worlds.

\end{document}